\documentclass[aps,prl,showpacs,floatfix,twocolumn,longbibliography]{revtex4-1}
\usepackage{mathrsfs}
\usepackage[figuresright]{rotating}
\usepackage{amsmath}
\usepackage{amssymb}
\usepackage{graphicx}
\usepackage{color}
\usepackage{dcolumn}
\usepackage{bm}
\usepackage[breaklinks=true,colorlinks=true,linkcolor=blue,urlcolor=blue,citecolor=blue]{hyperref}
\usepackage{ulem}
\usepackage{soul}

\def\sro{Sr$_2$RuO$_4$}

\def\o{$^{17}$O}

\def\be{\begin{equation}}
\def\ee{\end{equation}}

\makeatletter

\newcommand{\Rmnum}[1]{\expandafter\@slowromancap\romannumeral #1@}
\makeatother

\begin{document}

\title{Normal state $^{17}$O NMR studies of Sr$_{2}$RuO$_{4}$ under uniaxial stress}

\author{Yongkang Luo$^{1,2}$}
\email[]{mpzslyk@gmail.com}
\author{A. Pustogow$^{1}$}
\author{P. Guzman$^{1}$}
\author{A. P. Dioguardi$^{3}$}
\author{S. M. Thomas$^{3}$}
\author{F. Ronning$^{3}$}
\author{N. Kikugawa$^{4}$}
\author{D. A. Sokolov$^{5}$}
\author{F. Jerzembeck$^5$}
\author{A. P. Mackenzie$^{5,6}$}
\author{C. W. Hicks$^{5}$}
\author{E. D. Bauer$^{3}$}
\author{I. I. Mazin$^{7}$}
\author{S. E. Brown$^{1}$}
\email[]{brown@physics.ucla.edu}
\address{$^1$Department of Physics $\&$ Astronomy, UCLA, Los Angeles, CA 90095, USA;}
\address{$^2$Wuhan National High Magnetic Field Center and School of Physics, Huazhong University of Science and Technology, Wuhan 430074, China;}
\address{$^3$Los Alamos National Laboratory, Los Alamos, New Mexico 87545, USA;}
\address{$^4$National Institute for Materials Science, Tsukuba 305-0003, Japan;}
\address{$^5$Max Planck Institute for Chemical Physics of Solids, Dresden 01187, Germany;}
\address{$^6$Scottish Universities Physics Alliance, School of Physics and Astronomy, University of St Andrews, North Haugh, St Andrews KY16 9SS, UK;}
\address{$^7$Code 6393, Naval Research Laboratory, Washington, DC 20375, USA.}

\date{\today}

\begin{abstract}

The effects of uniaxial compressive stress on the normal state $^{17}$O nuclear magnetic resonance properties of the unconventional superconductor Sr$_{2}$RuO$_{4}$ are reported. The paramagnetic shifts of both planar and
apical oxygen sites show pronounced anomalies near the nominal $\mathbf{a}$-axis strain $\varepsilon_{aa}\equiv\varepsilon_v$, that maximizes the superconducting transition temperature, $T_{c}$. The spin susceptibility weakly increases on lowering the temperature below $T$$\simeq$10 K, consistent with an enhanced density of states associated with passing the Fermi energy through a van Hove singularity. Although such a Lifshitz transition occurs in the $\gamma$ band, formed by the Ru $d_{xy}$ states hybridized with in-plane O $p_{\pi}$ orbitals, the large Hund's coupling renormalizes the uniform spin susceptibilty, which, in turn, affects the hyperfine fields of all nuclei. We estimate this \textquotedblleft Stoner\textquotedblright\ renormalization, $S,$ by combining the data with first-principles calculations and conclude that this is an important part of the strain effect, with implications for superconductivity.

\end{abstract}
\maketitle

\section{\Rmnum{1}. Introduction}

The physics of the unconventional superconductivity (SC) of Sr$_{2}$RuO$_{4}$ \cite{Maeno:1994} remains a subject of longstanding importance, with particular focus on order-parameter symmetry \cite{Rice:1995}. There are numerous experimental results consistent with a chiral $p$-wave superconducting state \cite{Mackenzie:2003,Maeno:2012,Kallin:2012,Mackenzie:2017}, including evidence for time-reversal symmetry breaking for $T<T_{c}$ \cite{Luke:1998,Xia:2006}, and lack of suppression of the in-plane spin susceptibility on cooling through the superconducting critical temperature $T_c$, as deduced from nuclear magnetic resonance (NMR) spectroscopy \cite{Ishida:1998,Murakawa:2004} and neutron scattering \cite{Duffy:2000}. At the same time, there are other experimental results inconsistent with that interpretation \cite{Kirtley:2007,Hicks:2010,Yonezawa:2013,Hassinger:2017,Kittaka:2018}, and the out-of-plane spin susceptibility also remains constant \cite{Murakawa:2004}, in contradiction with the expectations for the chiral state \cite{Kallin:2012,Kim:2017}.

For several reasons, the normal-state physics of Sr$_{2}$RuO$_{4}$\ is equally
topical. It was anticipated at a very early stage that electron-electron interactions are controlled by the Hund's rule coupling \cite{Mravlje:2011}, and it was later shown within the dynamical mean field theory that the electrons are subject to strong Hund's rule correlations while the system remains metallic and far from the Mott insulator regime \cite{Medici:2011,Georges:2013}. Mean-field density functional theory (DFT) calculations within the Generalized Gradient Approximation (GGA) are unstable against ferromagnetism \cite{Kim:2017}. Even though strong correlations lead to fluctuations suppressing this instability, there still remains a substantial Stoner renormalization of the uniform spin susceptibility. This led to the analogy with the triplet superfluidity of $^{3}$He \cite{Mazin:1997}, anticipated earlier on the grounds that a related compound, SrRuO$_{3}$ is ferromagnetic \cite{Rice:1995}. Although later it was found that the leading magnetic instability is at a non-zero momentum $\mathbf{q_{0}}$$\approx$($\pm$0.6, $\pm$0.6, 0)$\pi/a$ \cite{Sidis:1999,Mazin:1999}, the proximity to a ferromagnetic state dominates the debate related to the superconducting order parameter symmetry \cite{Mackenzie:2017,Steffens:2018}.

An additional feature is the proximity to a 2D Lifshitz point \cite{Damascelli:2000} associated with a van Hove singularity (vHs), and the question as to its relationship to both normal state properties and nature of the superconducting state. Recently, striking physical property changes, including a factor of 2.5 increase in superconducting critical temperature $T_{c}$, from 1.4 K to 3.5 K \cite{Steppke:2017}, accompanied by a pronounced non-Fermi Liquid behavior of the resistivity \cite{Barber:2018}, were observed under application of in-plane strain $\varepsilon_{aa}$. This was tentatively interpreted as a Fermi level crossing of the vHs when $\varepsilon_{aa}$ reaches a critical value $\varepsilon_v$. Since direct experimental evidence is still lacking, it is important to test this interpretation in complementary studies of the normal state while subject to strain. Also, the vHs is expected to influence quite differently the triplet and singlet superconducting states, and this provides further motivation for physical property studies under strain. For singlet pairing, the order parameter (SC gap) can be large at the vHs (\textit{e.g., }for the $d_{x^{2}-y^{2}}$ symmetry), and thus the local density of states (DOS) enhancement at the vHs is very beneficial. On the contrary, the triplet order parameter at precisely the Lifshitz point is zero by symmetry, and therefore a triplet state is less suited to take advantage of the vHs unless the pairing interaction itself is enhanced. Since the DOS enhancement brings the system closer to ferromagnetism, the latter case is possible \cite{Kivelson:2018}.

With these issues in mind, we set out to verify experimentally that the same strain at which $T_{c}$ peaks indeed corresponds to a maximum in DOS, and to assess, as quantitatively as possible, the change in DOS and Stoner enhancements to the susceptibility under strained conditions. To this end, NMR measurements inform on the details of the normal state, through site and orbitally specific hyperfine couplings. Indeed, the enhancement is evident in the results presented, and moreover, the inferred enhancement semi-quantitatively accounts for the transport results in Ref. \cite{Barber:2018}. Looking ahead, it is worth emphasizing that the method is considered a litmus test for the superconducting state parity \cite{Ishida:1998,Imai:1998}, including any strain-induced order-parameter changes. The results presented in the next sections are normal state $^{17}$O\ NMR spectroscopy for in-plane $\mathbf{B}\parallel\mathbf{b}$, and out-of-plane $\mathbf{B}\parallel\mathbf{c}$ fields, as well as $^{17}$O NMR relaxation rates for $\mathbf{B}\parallel\mathbf{b}$ in the presence of $\mathbf{a}$-axis strain $\varepsilon_{aa}$. These are interpreted by way of complementary DFT calculations.

\section{\Rmnum{2}. Experimental details}

Single crystalline Sr$_{2}$RuO$_{4}$ used for these measurements was grown by the floating-zone method \cite{Maeno:1994}. Smaller pieces were cut and polished along crystallographic axes with typical dimensions 3$\times $0.3$\times$0.15 mm$^{3}$, and with the longest dimension aligned with the $\mathbf{a}$-axis. $^{17}$O isotope ($^{17}I$=5/2, gyromagnetic ratio $^{17}\gamma$=$-$5.7719 MHz/T \cite{Harris:2001}) spin-labelling was achieved by annealing in 50\% $^{17}$O-enriched oxygen atmosphere at 1050 $^{\circ}$C for 2 weeks \cite{Ishida:1998,AnnealNote}. The sample quality was not observably changed following this procedure, with $T_{c}$$\approx$1.44 K identified by specific heat measurements (Supplemental Materials, SM) \cite{SuppMat}. For the NMR experiments, the sample was mounted on a piezoelectric strain cell (Razorbill, UK) with an effective (exposed) length $L_{0}$$\sim$1 mm (see Fig.~S1a, SM). Three samples (labeled as S1, S2 and S3) were measured in this work. A nominal compressive stress is applied along the $\mathbf{a}$-axis, with corresponding strains ($\varepsilon_{aa}$$\equiv$$\delta L/L_{0}$) estimated to be up to $\sim-0.72$\% using a pre-calibrated capacitive dilatometer; the accuracy is limited by the unknown deformations of the epoxy clamp \cite{StrainMeasurement}. For reference, the observed maximum $T_c(\varepsilon_{aa})$ occurs at a quantitatively similar displacement as reported in Ref. \cite{Steppke:2017}, $T_c^{max}$=$T_c(\varepsilon_v)$, with $\varepsilon_v$$\simeq$$-0.6\%$. Most of the NMR measurements were performed at fixed temperature $T$=4.30(5) K and carrier frequency $f_{0}$$=$46.8 MHz ($B\simeq$8.1 T), using a standard Hahn echo sequence. Spectra, including satellite transitions, were collected in field-sweep mode, whereas a close examination of the central transition (-1/2$\leftrightarrow$1/2) for both in-plane and apical sites was carried out under fixed-field conditions. Some field and temperature dependence was explored, too. The application of NMR in conjunction with the piezoelectric-driven \textit{in situ} strain is particularly challenging, because of the severe constraints on sample size. As a result, some modifications to standard resonant tank circuit configurations were adopted.

For insight into the strain-induced changes to the NMR shifts, and particularly those associated with the vHs, Density Functional calculations using the Linear Augmented Plane Wave package WIEN2k \cite{Blaha:2001} were performed, including spin-orbit interaction. The specific objective was to extract at least semi-quantitative information about the origin, evolution, and relative importance of the various individual contributions to the net Knight shifts. A local density approximation (LDA) for the exchange-correlation functional, a $k$-point mesh of 41$\times$41$\times$41, and the expansion parameter $RK_{\max}$$=$$7$ were utilized. Further, the optimized structures of Ref.~\cite{Steppke:2017} were used, and then interpolated to assure that the strain at which the vHs crosses the Fermi level is included. It turns out that the proximity to a (ferromagnetic) quantum critical point forced some adjustments to the standard procedure. One reason lies with the mean-field approach: DFT overestimates the tendency to magnetism, because in reality the Hund's rule derived interaction $I$ and, correspondingly, the Stoner renormalization $S$, are reduced by quantum fluctuations that are not accounted for. A second challenge originates with the very narrow calculated DOS singularity at the vHs: in relation to the Knight shift evaluation, an external magnetic field is applied followed by the computation of generated hyperfine fields. The singularity full width at half maximum is $\sim$3 meV, and holds only 0.0015 e$^{-}$ in each spin channel. As a result, an external field producing sufficiently strong hyperfine fields (compared to the computational noise), is too large to properly monitor the vHs peak. Nevertheless, the calculations at the larger fields produce useful information, in part because the origin of the net Knight shifts in terms of individual contributions is obtained.

\begin{figure}[t]
\includegraphics[width=\columnwidth]{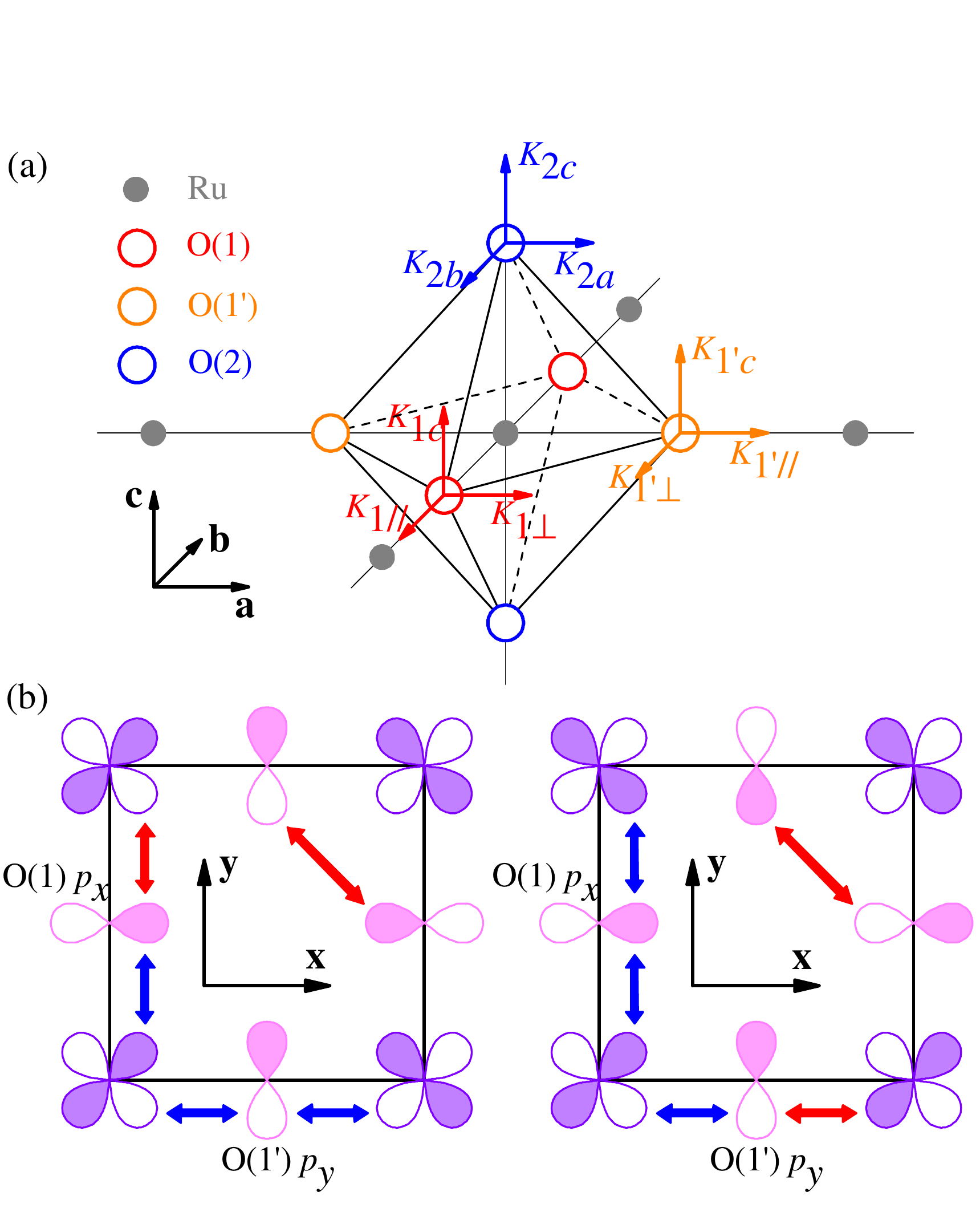}\caption{(a)
Configurations of planar O(1) and O(1$^{\prime}$) in RuO$_{2}$ plane and
apical O(2) in the SrO layer around Ru ion in a unit cell of Sr$_{2}$RuO$_{4}%
$. Compressive strain is applied along the $\mathbf{a}$-axis ($\varepsilon
_{aa}$); magnetic fields are applied orthogonal, $\parallel\mathbf{b},
\parallel\mathbf{c}$. Arrows signify the principal axes of Knight shift
tensors. (b) Orbitals forming the $\gamma$ band at the X (left) and Y (right)
points in the Brillouin zone. The blue (red) double-arrows show positive
(negative) signs of orbital overlaps. Note that at the Y point only
O(1)$p_{x}$ orbitals participate in the band formation, while O(1$'$)$p_{y}$
suffers from cancellation of the left and right overlaps. The
weak O(1)-O(1$'$) overlaps also cancel out, as shown in the figure.}
\label{fig:SiteGeometry}%
\end{figure}

\begin{figure}[th]
\includegraphics[width=\columnwidth]{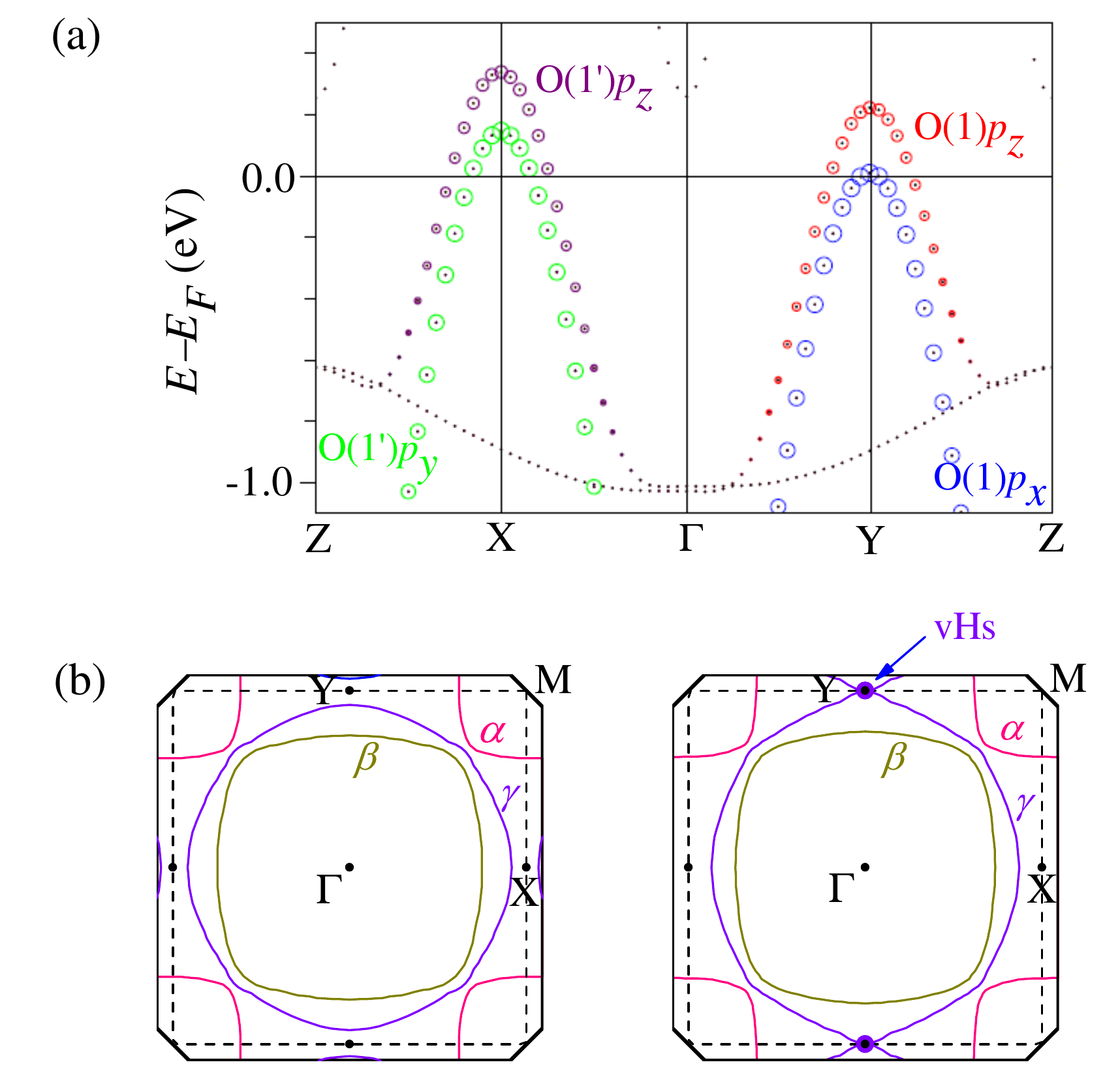}
\caption{(a) Bands along the $\Gamma$-X and $\Gamma$-Y directions. The partial
weights of the O(1)$p_{x}$, O(1$^{\prime}$)$p_{y}$, O(1)$p_{z}$, and
O(1$^{\prime}$)$p_{z}$ orbitals are shown in green, blue, red and purple,
respectively. Other oxygen orbitals have far lesser weight near the Fermi energy. (b)
Depictions of the 2D Fermi surfaces, with quasi-2D $\gamma$ ($d_{xy}$) and q1D
$\alpha, \beta$ ($d_{zx,yz}$) bands.}%
\label{fig:BandsFermiSurface}%
\end{figure}

\section{\Rmnum{3}. Results and Discussion}

The geometry of our experiment is depicted in Fig.~\ref{fig:SiteGeometry}(a) \cite{Mukuda:1998}. Each Ru ion is coordinated octahedrally by four planar O(1) and two apical O(2) oxygen sites, with a small elongation along the $\mathbf{c}$-axis. While $\mathbf{a}$-axis strain $\varepsilon_{aa}$ renders the sites O(1) and O(1$^{\prime}$) crystallographically inequivalent, their local symmetries are different even for the unstrained case, and external field $\mathbf{B}\parallel\mathbf{b}$. The field-sweep spectra in Fig.~S2 are described by parameters (shifts, electric field gradient (EFG)) similar to previous reports for $^{17}$O\ NMR in the unstrained Sr$_{2}$RuO$_{4}$\ \cite{Mukuda:1998,Imai:1998}, with five NMR transitions for each of 3(2) distinct sites for $\mathbf{B}\parallel\mathbf{b}$ ($\mathbf{B}\parallel\mathbf{c}$) \cite{SuppMat}.

The most relevant orbitals for the \o\ couplings are Ru 4$d$ $t_{2g}$, which hybridize with O $p$ states to form the quasi-2D $\gamma$ band, predominantly from the $d_{xy}$ orbital, and similarly the quasi-1D $\alpha$ and $\beta$ bands from the $d_{zx,yz}$ orbitals, Fig. \ref{fig:SiteGeometry}b. The spin-orbit coupling (SOC) mixes these. While mixing is strongest along the Brillouin zone diagonal ($\Gamma$$-$$\text{M}$ in momentum space, see Fig.~\ref{fig:BandsFermiSurface}b) \cite{Pavarini:2006,Haverkort:2008}, it is more important here that it mixes the $d_{xy}$ and $d_{yz}$ bands at Y. The latter has the effect of pushing down the lower band ($d_{xy}$) by about 20 meV, which shifts the critical strain $\varepsilon_{v}$ where the Lifshitz transition shows up in the calculations, from about $\sim-$1.0\% to $\sim-0.85$\%. Additional mass renormalization, not accounted for in the DFT calculations, reduces the critical strain still further, consistent with the experimentally observed maximum in $T_{c}$ between $-0.55$\% and $-0.60$\% \cite{Steppke:2017,SuppMat}.

\begin{figure}[t]
\vspace*{5pt} \hspace*{-7pt}
\includegraphics[width=9.0cm]{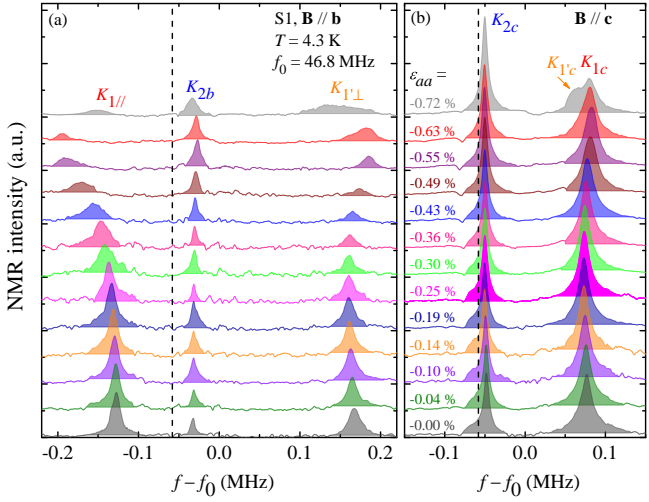}
\vspace*{-10pt}
\caption{NMR spectra of the central transitions ($\frac{1}{2}\leftrightarrow-\frac{1}{2}$) of O(1), O(1$^{\prime}$) and O(2) at various strains for magnetic field along $\mathbf{b}$- (left), $\mathbf{c}$-axes (right). The measurements were carried out at fixed temperature ($T$=4.3 K) and field ($B$=8.0970 T) and radio frequency $f_{0}$=46.80 MHz. The curves are vertically offset for clarity. The dash vertical line
corresponds to $^{17}\gamma$=$-$5.7719 MHz/T (D$_{2}^{17}$O) \cite{Harris:2001} with zero shift.}
\label{fig:CentralTransitionsStrain}%
\end{figure}

\begin{figure}[t]
\includegraphics[width=\columnwidth]{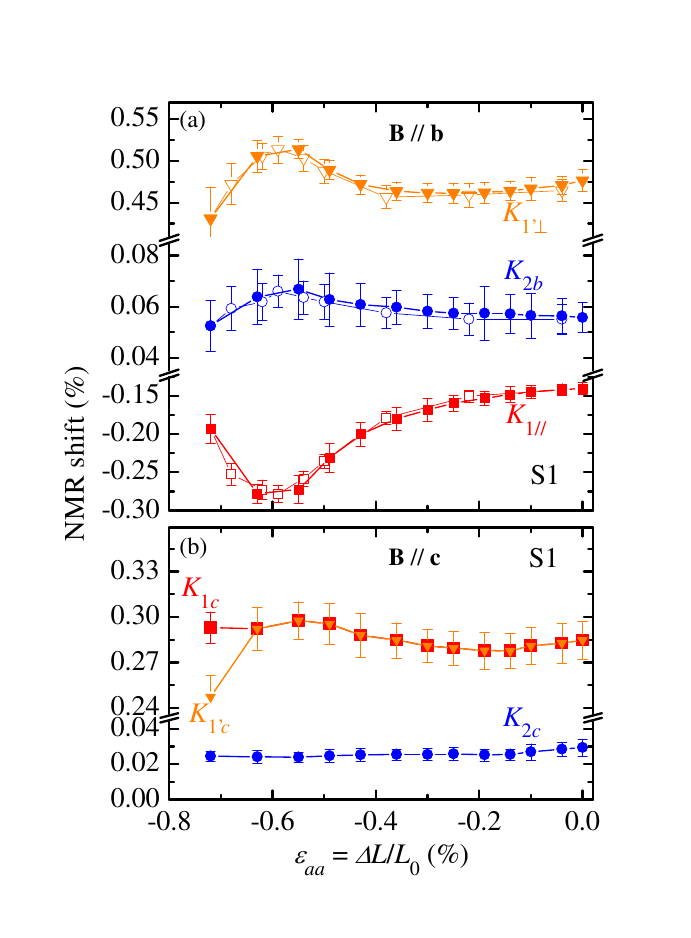}
\caption{Measured NMR shifts for $\mathbf{B}\parallel\mathbf{b}$ (a) and for $\mathbf{B}\parallel\mathbf{c}$ (b) at $T$=4.3 K. The solid (open) symbols represent for increasing (decreasing) $|\varepsilon_{aa}|$. The error bars are determined by the half width of the peaks. Similar results were reproduced from several samples, see Fig.~S5a.}
\label{fig:shifts}%
\end{figure}

$^{17}$O NMR spectra under varying strain conditions are shown in Fig. \ref{fig:CentralTransitionsStrain}. The two panels depict the central transition for all three sites O(1), O(1$^{\prime}$), O(2), measured with carrier frequency $f_{0}$=46.80 MHz and magnetic field $B$=8.0970 T, applied parallel to $\mathbf{b}$ (left panel) and $\mathbf{c}$ axes (right panel), respectively. The peaks for O(1), O(1$^{\prime}$), O(2) appear at the labelled frequencies, measured relative to $f_0$. The vertical dashed lines correspond to zero shift. O(2), having relatively minor contribution to the Ru bands (there is only a weak coupling of the O(2) $p_{x,y}$ with Ru $d_{zx,yz}$ orbitals, respectively) exhibits a very small Knight shift. In contrast, Knight shifts for O(1) and O(1$^{\prime}$) vary strongly and show clear extrema at strain $\varepsilon_{aa}$=$\varepsilon_v$, corresponding to the putative vHs and defined as where $T_{c}(\varepsilon_{aa})$ is largest. The anomaly is most pronounced for the in-plane field orientation. For larger strains, there is significant broadening, tentatively attributed to a strong strain dependence of the spin susceptibility, combined with a distribution of strains within the sample. (Note that asymmetries in mounting geometry lead naturally to crystal bending.) In the right-hand panel, O(1,1$^{\prime}$) spectral peaks appear indistinguishable at small strain, with pronounced broadening and splitting appearing for strains exceeding $\varepsilon_{v}$. The NMR shifts $K$, defined as the relative change of resonance frequency referenced to that observed for D$_2$$^{17}$O, are shown as a function of $\varepsilon_{aa}$ in Fig.~\ref{fig:shifts}. Similar results reproduced from other samples can be found in Fig.~S5a. One striking feature is that the Knight-shift anomaly near $\varepsilon_{aa}$$\approx$$\varepsilon_{v}$ is seen in all the measured $^{17}$O sites, not only in O(1) that is most relevant to vHs at Y.

In metals, the NMR shift is governed by three main contributions resulting from spin and orbital responses to the applied field: (i) isotropic coupling from the Fermi contact interaction and core polarization, (ii) anisotropic coupling of the dipolar field generated by the electronic spin away from the nucleus, and (iii) fields generated by orbital currents. For computational purposes, this partitioning of the hyperfine field contributions can be summarized as
\begin{equation}
\mathbf{h(r)=h}_{s}+\mathbf{h}_{d}\mathbf{+h}_{o}=-\beta\left[  \frac
{8\pi\mathbf{s}\delta(\mathbf{r)}}{3}+\frac{3\mathbf{r(r\cdot s)-}%
r^{2}\mathbf{s}}{r^{5}}+\frac{\mathbf{L}}{r^{3}}\right]  , \label{hff}%
\end{equation}
where $\mathbf{s}$ is the spin moment of an electron, and \textbf{L }its orbital moment. Real space integration results in the total local field. Note that $\mathbf{h}_{s}$ has no anisotropy, while $\mathbf{h}_{d}$ gives no isotropic contribution ($\mathbf{h}_{o}$ has both).

\begin{figure}[ptb]
\includegraphics[width=\columnwidth,angle=0]{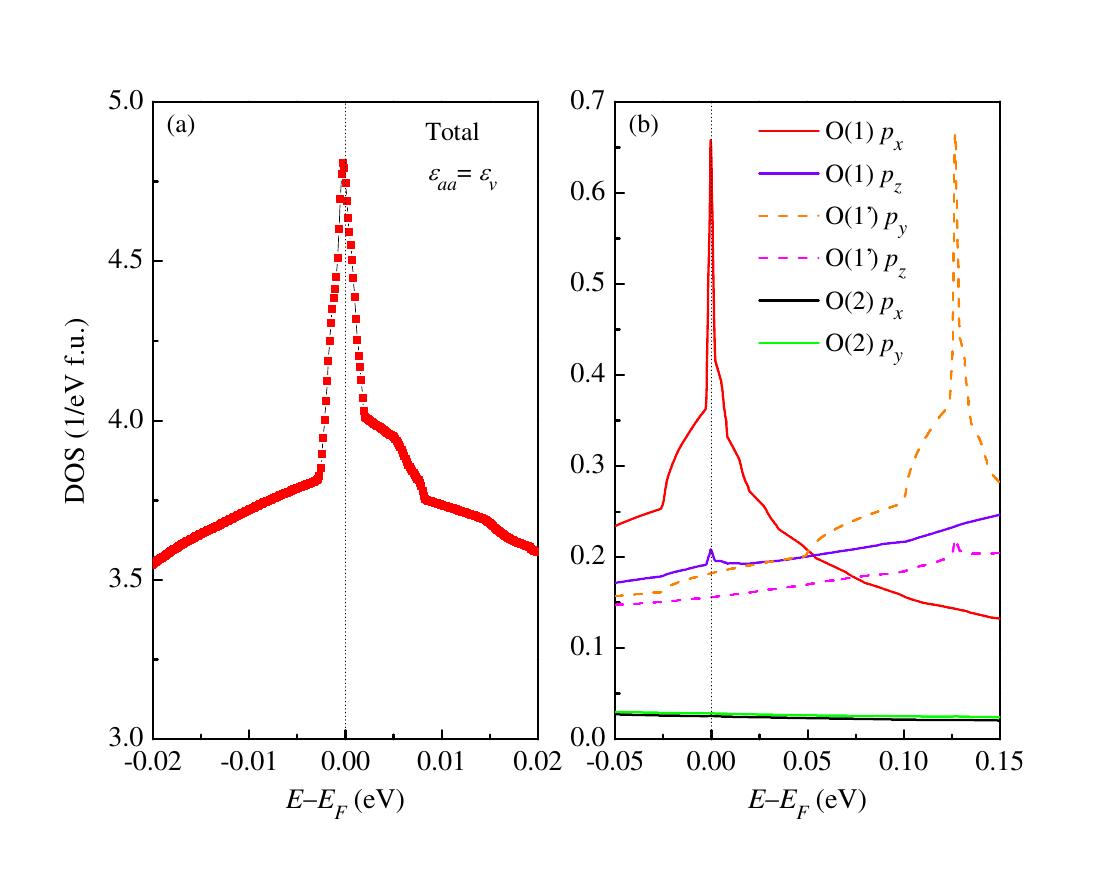}
\caption{(a) Calculated density of states (DOS) at the critical strain, at
which the calculated van Hove singularity is located exactly at the Fermi
level. Note the very small width (3 meV) and weight (0.0015 e$^-$ per spin channel)
of the peak in DOS. (b) Partial DOS projected onto different O orbitals. The
orbitals that are not shown have negligible weight.}%
\label{fig:DOS}%
\end{figure}

The net spin magnetization is written as
\begin{equation}
M_s=\chi_s H%
\end{equation}
where the full uniform spin susceptibility $\chi_s$ can be expressed using the Stoner factor $S$
\begin{equation}
\chi_s=(m^*/m_0) S \chi_{s0}^{DFT},%
\label{Eqchis}
\end{equation}
where $\chi_{s0}^{DFT}$ is the \textit{noninteracting} spin susceptibility proportional to (neglecting spin-orbit effects) the DFT density of states, and the factor of $m^*/m_0$ arises from mass renormalization beyond the scope of DFT. Writing $S$ in the Random-phase approximation (RPA) \cite{StonerTensor} guides our expectations for its evolution under strained conditions,
\begin{equation}
S_{RPA}=\frac{1}{1-IN(E_{F})},%
\label{S}
\end{equation}
where $N(E_F)$ is the actual DOS. Then, the total uniform magnetic field is the sum of the external field and the induced response, the latter being enhanced compared to the noninteracting case by the factor $S$. Note that the orbital moment $\mathbf{L}$ in Eq. (\ref{hff}) is assumed to be generated by the spin magnetization through spin-orbit coupling. In addition, there is another orbital term (paramagnetic van Vleck), which is not enhanced in the same way as $\chi_s$. While usually considered small \cite{Imai:1998}, an accurate accounting is not expected in the DFT framework. As indicated by Eqs. (\ref{Eqchis},\ref{S}), the strain-dependent enhancement of $S$ is important as a mechanism for transferring anomalous responses (linked to the vHs) to orbitals other than Ru $d_{xy}$ and the corresponding hybridizing O$p$. Notable also is that, in principle, $S$ can be more sensitive to the enhancement of the DOS than $\chi_{s}$ itself. To establish relevance, consider that inelastic neutron scattering measurements indicated $\chi_{s}$ is enhanced by about a factor of 7 compared to the DFT DOS, viz.
$\frac{\chi_s(\varepsilon_{aa}=0)}{\chi_{s0}^{DFT}(\varepsilon_{aa}=0)}\sim7$ \cite{Steffens:2018}, with the enhancement originating from a mass renormalization factor ($m^{\ast}/m_{0}\sim3.5$ \cite{Mackenzie:2003}), and an inferred Stoner factor ($S\sim 2$). Using Eq.~(\ref{S}), $IN(E_{F})\approx0.5$ at zero strain, and with $N(E_{F})$ increased by 30\%, as in Fig.~\ref{fig:DOS}a, leads to an inferred increase of $S$ from 2 to 3. Thus, if $m^{\ast}/m_{0}$ and $I$ are taken as strain-independent, one gets $\frac{\chi_s(\varepsilon_{aa}=\varepsilon_v)}{\chi_{s0}^{DFT}(\varepsilon_{aa}=\varepsilon_v)}\sim10.5$, $\frac{\chi_{s}(\varepsilon_{aa}=\varepsilon_v)}{\chi_{s0}^{DFT}(\varepsilon_{aa}=0)}\sim14$, and thus $\frac{\chi_s(\varepsilon_{aa}=\varepsilon_{v})}{\chi_s(\varepsilon_{aa}=0)}\sim 2$, namely a factor of 2 enhancement in actual spin susceptibility at the critical strain relative to zero strain.

Symmetry considerations indicate that only O(1)$p_{x}$ orbitals couple with Ru$d_{xy}$ states at Y, and therefore, only the O(1)$p_{x}$ orbitals are expected to be directly sensitive to the vHs (see Fig.~\ref{fig:DOS}b). Thus, one might infer that only the O(1) Knight shift should be affected by the DOS peak at the vHs. However, on general grounds, all sites are sensitive, because of the increased Stoner enhancement factor $S$. Indeed, all measured Knight shifts are affected by strain (Fig. \ref{fig:shifts}), with $K_{1\parallel}$ more so, presumably because of the direct influence of increased $\gamma$ DOS. The strain-induced reduction of the Korringa ratio\cite{Korringa:1950,Hirata-ModifiedKorringa}, shown in the inset of Fig.~S5a for the case $\mathbf{B}\parallel\mathbf{b}$ \cite{SuppMat}, is consistent with an enhanced Stoner factor $S$.

\begin{figure}[ptb]
\includegraphics[width=\columnwidth]{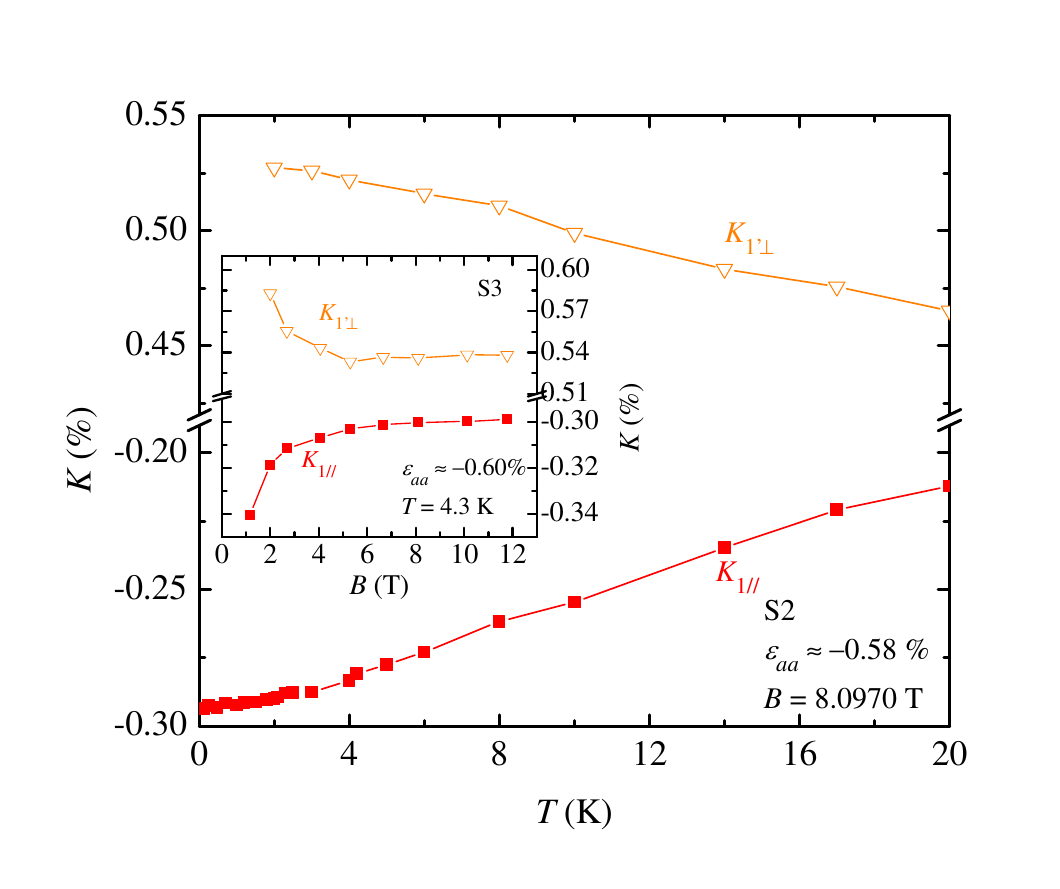}
\caption{Main panel, temperature dependence of $K_{1\parallel}$ and $K_{1^{\prime}\perp}$,
evaluated at the critical strain $\varepsilon_{v}$, $B$ = 8.0970 T, and
$\mathbf{B}\parallel\mathbf{b}$. Inset, field dependence of $K_{1\parallel}$
and $K_{1^{\prime}\perp}$ measured at $\varepsilon_{v}$ and 4.3 K.}%
\label{fig:TBdependences}%
\end{figure}

Experimental evidence for the narrow vHs and its influences on physical properties is shown in Fig.~\ref{fig:TBdependences}, which depicts shifts with strikingly strong field and temperature dependences for $\varepsilon_{aa}=\varepsilon_v$. The observations are qualitatively consistent with comparable energy scales for Zeeman, thermal, and vHs terms, where, for instance, the broadening of the Fermi distribution progressively weakens the sensitivity of thermodynamic properties to the vHs, even when it is situated precisely at the chemical potential\cite{Nourafkan:2018}. Similar observations for the magnetization were previously reported in a doping study, in which the effects of substitution of La for Sr in Sr$_{2-y}$La$_y$RuO$_4$ were interpreted as evidence for moving $\gamma$-band Fermi energy to the X and Y points of the Brillouin zone \cite{Kikugawa:2004}. These behaviors are even more striking when compared to expectations in a single-particle framework, because the Zeeman coupling shifts the vHs singularity away from the chemical potential. The saturating temperature-dependence at fixed field strength that follows is at odds with observations, and warrants further study in the context of quantum critical behavior which can be boosted by Stoner enhancement (see below).

\begin{figure}[t]
\includegraphics[width=8.5cm]{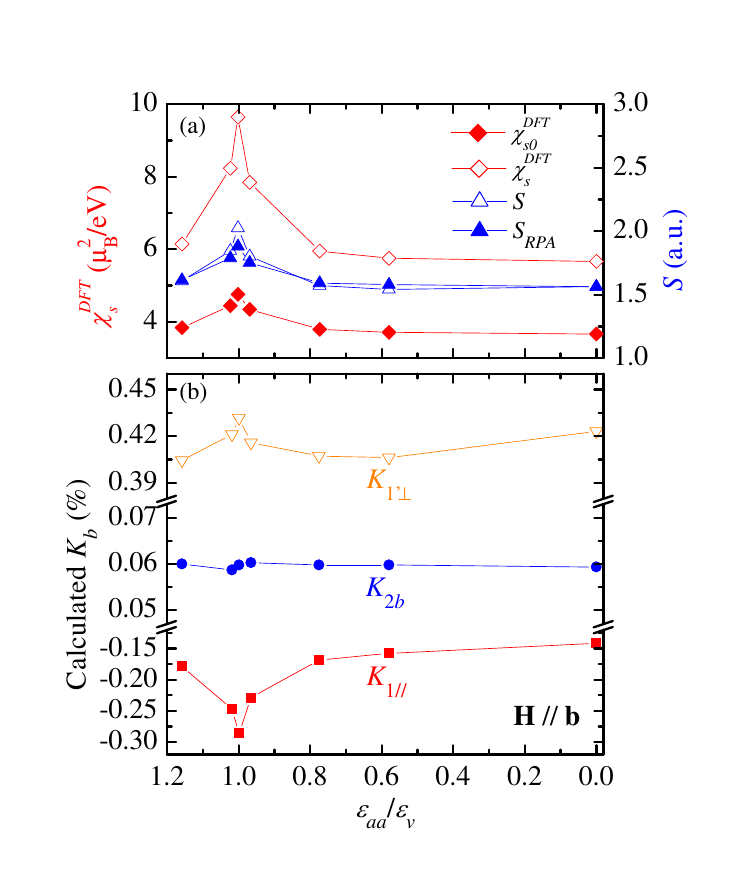}
\caption{(a) Calculated magnetic susceptibility in DFT. $\chi_{s0}^{DFT}\equiv\mu_{B}^2 N(E_F)$ is the noninteracting susceptibility, $\chi_s^{DFT}$ is obtained by dividing the calculated magnetization by the applied field, $M_s/\mu_{B}H$. The average DFT Stoner factor $S=\chi_s^{DFT}/\chi_{s0}^{DFT}$, and $S_{RPA}=1/[1-IN(E_F)]$. Here, $S_{RPA}$ is normalized to $S$ obtained from the calculated DFT result at zero strain. Its variation with strain is calculated from Eq.~\ref{S}, and the strain dependent DFT density of states. (b) Calculated total Knight shifts for $\mathbf{H}\parallel\mathbf{b}$ for the three sites, O(1), O(1$'$) and O(2), as a function of normalized strain. See the text for details.}%
\label{chi}%
\end{figure}

For a semiquantitative evaluation of the Stoner enhancement and the subsequent impact on the observable quantities, the data were contrasted to the results of the DFT calculations. As stated, the inherent deficiency of the DFT calculations for such a strongly correlated material as \sro\ forced deviations from the usual procedure. The standard calculations, such as those presented in Ref.~\cite{Steppke:2017}, are unstable against spontaneous formation of a ferromagnetic state. The tendency toward this instability was reduced, somewhat arbitrarily, by scaling the Hund's coupling by half. This ensured numerically stable calculations in external fields up to at least 5 T, even at $\varepsilon_{aa}$=$\varepsilon_{v}$. The impact of the reduced Hunds' coupling appears to produce systematic errors in related absolute parameters, but less so for the relative changes induced by strain. For example, for the selected scaling, Fig.~\ref{chi}a indicates that the calculated $\chi_s^{DFT}(\varepsilon_{aa}$=$0)$ renormalization is $\sim$1.6, whereas the known correlation-induced mass enhancement is about 3.5 \cite{Mackenzie:2003}. Therefore, the downscaling is too strong. Given this caveat, at the critical strain, $\chi_s^{DFT}(\varepsilon_{aa}$=$\varepsilon_{v})$ is enhanced over $\chi_s^{DFT}(\varepsilon_{aa}$=$0)$ by about 70\%, while $S(\varepsilon_v)$ itself is enhanced by a much smaller factor, about 30\% over $S(0)$ (right frame of Fig.~\ref{chi}a). The scaling for the shifts should follow approximately these factors. Namely, the enhancement of $K_{1\parallel}$ is expected to scale with $\chi_s$, and therefore of order 70\%, whereas the enhancement of $K_{1^{\prime}\perp}$, being sensitive to enhancement of $S$, is expected to be much smaller, of order 30\%. The former matches well to the data in Fig.~\ref{fig:shifts}, as well as the calculations presented in Fig.~\ref{chi}. The latter enhancement of 30\% is relatively larger than the experimental results (Fig.~\ref{fig:shifts}a), as well as the calculations (Fig. \ref{chi}b), which are both $\simeq10\%$. The discrepancy could be associated with unaccounted-for nonsingular contributions, such as in an orbital part (van Vleck or induced through spin-orbit coupling), or nonlinearities, as documented in Fig.~\ref{fig:TBdependences}.

Therefore, the qualitative conclusions from the experiments and in comparing to the DFT calculations are as follows: (1) there are two mechanisms for enhancing the Knight shifts near the critical strain, one applicable to all sites and field directions, and the other only to $K_{1\parallel}$. Both are directly related to the DOS enhancement and show unambiguously that the maximum in $T_{c}$ indeed coincides with that in DOS; (2) ferromagnetic spin fluctuations intensify substantially at the same strain due to Stoner enhancement. This effect may also play a key role in boosting $T_{c}$; (3) the nonlinear magnetic response for $\varepsilon_{aa}\simeq\varepsilon_v$ and at low temperatures and magnetic fields appears to deviate from the expected single particle response, and are offered here as evidence for both the enhancement of the spin fluctuations, as well as the proximity to a ferromagnetic instability.

Expanding further on point (3) above, the strain-dependent enhancement of $S$ provides a natural explanation to the recently reported resistivity measurements on stressed samples\cite{Barber:2018}, in which deviations from standard Fermi liquid behavior were observed and interpreted in terms of the DOS singularity\cite{Hlubina:1996}. In fact, the behavior may also be connected to the enhanced Stoner factor near the critical strain. Reported was the existence of a crossover temperature $T^*$, at which the electrical resistivity $\rho$=$\rho_0$$+$$AT^{\delta}$ changes from the Fermi-liquid behavior $\delta=2$ to $\sim$1.5-1.6. Note that this is close to what is expected for ferromagnetic spin-fluctuation behavior ($\delta$$\sim$4/3-5/3) \cite{Stewart:2001}. Moreover, $T^*\varpropto S^{-1}$ varies strongly with strain (see Fig.~S1b \cite{SuppMat}), and is minimized at $\varepsilon_v$. Both this observation, and the nonlinearities in the shifts (Fig. \ref{fig:TBdependences}) indicate $S$ peaks at $\varepsilon_v$.

Finally, some comments on the data collected for field aligned parallel the $\mathbf{c}$-axis are in order. In principle, one would expect similar behavior to that for the in-plane field, however, it appears that $K_{c}$s behave in a way difficult to rationalize in total. For strain $\varepsilon_{aa}\le-0.63\%$, a single absorption peak at $\sim0.29\%$ shift is observed for O(1,1$'$), with only a small increase in the range of $\varepsilon_{v}$. For larger strain, $\varepsilon_{aa}=-0.72\%$, the peak broadens considerably, and could be construed as exhibiting two components, but with drastically reduced first moment. The drop in intensity is likely a $T_1$ effect, a consequence of a (relatively) rapid pulse repetition rate (see Fig.~S5b, SM). The apparent spectral line ``splitting'' and distorted lineshape are consistent with what could result from a strain gradient along with a nonlinear variation of shift with strain. The main challenge, however, is to explain the observed evolution on approaching $\varepsilon_{v}$ from smaller strain, where the DFT calculations indicate larger shifts for O(1) than for O(1$'$).

It is possible that the orbital contributions play a more important role for this field orientation ($\mathbf{B}\parallel\mathbf{c}$). Interestingly, for the orbital part of $K_{1c}$, and to some extent, of $K_{1^{\prime}c},$ the calculations predict a sizeable enhancement, suggesting that the van Vleck contribution is not dominant, or, at least, less prominent here than for the in-plane fields, and, conversely, the SOC induces sizeable orbital Knight shifts. Moreover, the sign of this orbital contribution is opposite to the spin shifts, so there is a tendency toward cancellation. It is believed that correlation effects enhance the SOC in Sr$_{2}$RuO$_{4}$ by about a factor of two \cite{Kim:2018}. Empirically, if the O(1) and O(1$^{\prime}$) shifts are assumed to be entirely generated by SOC, while the O(2) shift is entirely van Vleck, a reasonable agreement with experiment is obtained, but with small but not negligible peak splittings for strains near $\varepsilon_{v}$ (Fig.~S6). Clearly, the NMR spectra for the field parallel to $\mathbf{c}$ require further investigations.

\section{\Rmnum{5}. Conclusion}

It is demonstrated, by means of the NMR spectroscopy under uniaxial stress, and corresponding density functional calculations, that there are \textit{two} different effects associated with the strain-induced vHs, which both need to be taken into account, namely the enhancement of the DOS associated with the $\gamma$-band Fermi energy passing through the vHs at the Y point of the Brillouin zone, and a substantial Stoner enhancement $S$. Associated with the enhanced $S$ is an intensification of ferromagnetic spin fluctuations and strong nonlinearities in the spin susceptibility to the lowest temperatures studied. This finding has immediate ramifications for superconductivity. Namely, first, the DOS is enhanced near the vHs point. In the first approximation, this effect strongly favors some singlet pairings, such as extended $s$, $d_{zx}\pm id_{yz}$ or $d_{x^2-y^2}$, mildly favors the $d_{xy}$ pairing, and less so any triplet pairing. However, this enhancement of the DOS, through the Stoner factor, boosts FM spin fluctuations, which favors a triplet states and would seem to \textit{disfavor} singlet pairing. Experimentally and theoretically, these two effects are comparable, and therefore it is unclear which is stronger. More information will be gained by studying NMR in the superconducting state as a function of strain.

\section{Acknowledgments}

We thank S. A. Kivelson, S. Raghu, J. D. Thompson, and F. Zhang for insightful discussions, and J. D. Thompson for magnetic properties characterization. This work was supported in part by the Laboratory Directed Research and Development (LDRD) program of Los Alamos National Laboratory under project number 20170204ER. Y.L. acknowledges partial support through the LDRD and 1000 Youth Talents Plan of China. N.K. acknowledges the support from JSPS KAKNHI (18K04715). I.I.M was supported by ONR through the NRL basic research program. This work was supported in part by the National Science Foundation (DMR-1410343 and DMR-1709304).

\section{Appendix: F\lowercase{urther consideration regarding} S\lowercase{toner renormalization}}

\begin{figure}[ptb]
\vspace{-50pt}
\includegraphics[width=1\columnwidth]{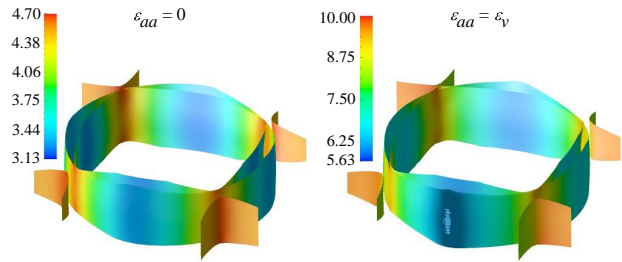}
\caption{\label{fig:ExchangeSplitting} Calculated Fermi surfaces (nonrelativistic) with no orthorhombic strain (left) and the strain corresponding to the vHs (right). No additional scaling of the Stoner interaction was applied, as opposed to the Knight shift calculation (Fig.~\ref{chi} and main text). The surfaces are colored with the calculated exchange splitting in a small uniform external field $H$, normalized to $\mu_B H$=1.6 meV. Note the different color scales for the two panels.}
\end{figure}

The experiments and calculations clearly demonstrate the importance of Stoner renormalization near the critical strain, but this is evaluated only semiquantitatively. For example, the RPA-like Eq.~(\ref{S}) implies a uniform renormalization of the exchange splitting over the entire Fermi surface. In actual calculations, this splitting varies substantially over the Fermi surface (depicted in Fig.~\ref{fig:ExchangeSplitting}). Still, it remains a qualitatively reasonable approximation. In Fig.~\ref{chi} we show the results of direct calculations of spin susceptibility, inferred by calculating the induced magnetization $M_{s0}(H)$ resulting from a small applied field. The full DFT susceptibility shown in Fig.~\ref{chi} is $M_s(H)/H$, and the Stoner factor $S=M_s(H)/M_{s0}(H)$, with $M_{s0}(H)=\mu_{B}^{2}N(E_F)H$ the Pauli result for non-interacting electrons.

Fig.~\ref{fig:ExchangeSplitting} indicates that the exchange splitting, for the same external field, is larger for the $\alpha$ and $\beta$ bands, than for the $\gamma$ band, and that this disparity is about twice larger at the critical strain than for the unstrained structure. Overall, in the former the local Stoner factor (the enhancement of the exchange splitting of the electronic states) varies between 3.2 and 4.7, and in the latter between 5.7 and 10.0, about a factor of two larger than for the unstrained structure. Consequently, it is entirely possible that this variation will weight differently the dipole and the spin-contact contributions. This is consistent with the fact that the temperature dependence of the in-plane, and only in-plane Knight shifts are opposite to that of the uniform susceptibility at $T\lesssim100$ K, and only these are affected by the vHs in our experiment\cite{Imai:1998}.


\newpage

\setcounter{table}{0}
\setcounter{figure}{0}
\setcounter{equation}{0}
\renewcommand{\thefigure}{S\arabic{figure}}
\renewcommand{\thetable}{S\arabic{table}}
\renewcommand{\theequation}{S\arabic{equation}}
\onecolumngrid

\newpage

\begin{center}
{\bf \large
{\it Supplemental Material:}\\
Normal state $^{17}$O NMR studies of Sr$_2$RuO$_4$ under uniaxial stress
}
\end{center}

\small
\begin{center}
Yongkang Luo$^{1,2*}$\email{mpzslyk@gmail.com}, A. Pustogow$^{1}$, P. Guzman$^{1}$, A. P. Dioguardi$^{3}$, S. M. Thomas$^{3}$, F. Ronning$^{3}$, N. Kikugawa$^{4}$, D. A. Sokolov$^{5}$, F. Jerzembeck$^5$, A. P. Mackenzie$^{5,6}$, C. W. Hicks$^{5}$, E. D. Bauer$^{3}$, I. I. Mazin$^{7}$, and S. E. Brown$^{1\dag}$\email{brown@physics.ucla.edu}\\
$^1${\it Department of Physics and Astronomy, University of California, Los Angeles, CA 90095, USA;}\\
$^2${\it Wuhan National High Magnetic Field Center and School of Physics, Huazhong University of Science and Technology, Wuhan 430074, China;}
$^3${\it Los Alamos National Laboratory, Los Alamos, New Mexico 87545, USA;}\\
$^4${\it National Institute for Materials Science, Tsukuba 305-0003, Japan;}\\
$^5${\it Max Planck Institute for Chemical Physics of Solids, Dresden 01187, Germany;}\\
$^6${\it Scottish Universities Physics Alliance, School of Physics and Astronomy, University of St Andrews, North Haugh, St Andrews KY16 9SS, UK; and}\\
$^7${\it Code 6393, Naval Research Laboratory, Washington, DC 20375, USA.}\\

\date{\today}
\end{center}

\normalsize
\vspace*{15pt}
In this \textbf{Supplemental Material (SM)}, we provide the experimental setup, sample characterization, field-swept spectra, spin-lattice relaxation rate and electric field gradient (EFG) results of Sr$_2$RuO$_4$ under uniaxial stress, as well as additional theoretical details that further support the discussions in the main text.\\

\section{\textbf{SM \Rmnum{1}: S\lowercase{ample characterizations}}}

\begin{figure*}[ht]
\vspace*{-20pt}
\includegraphics[width=16cm]{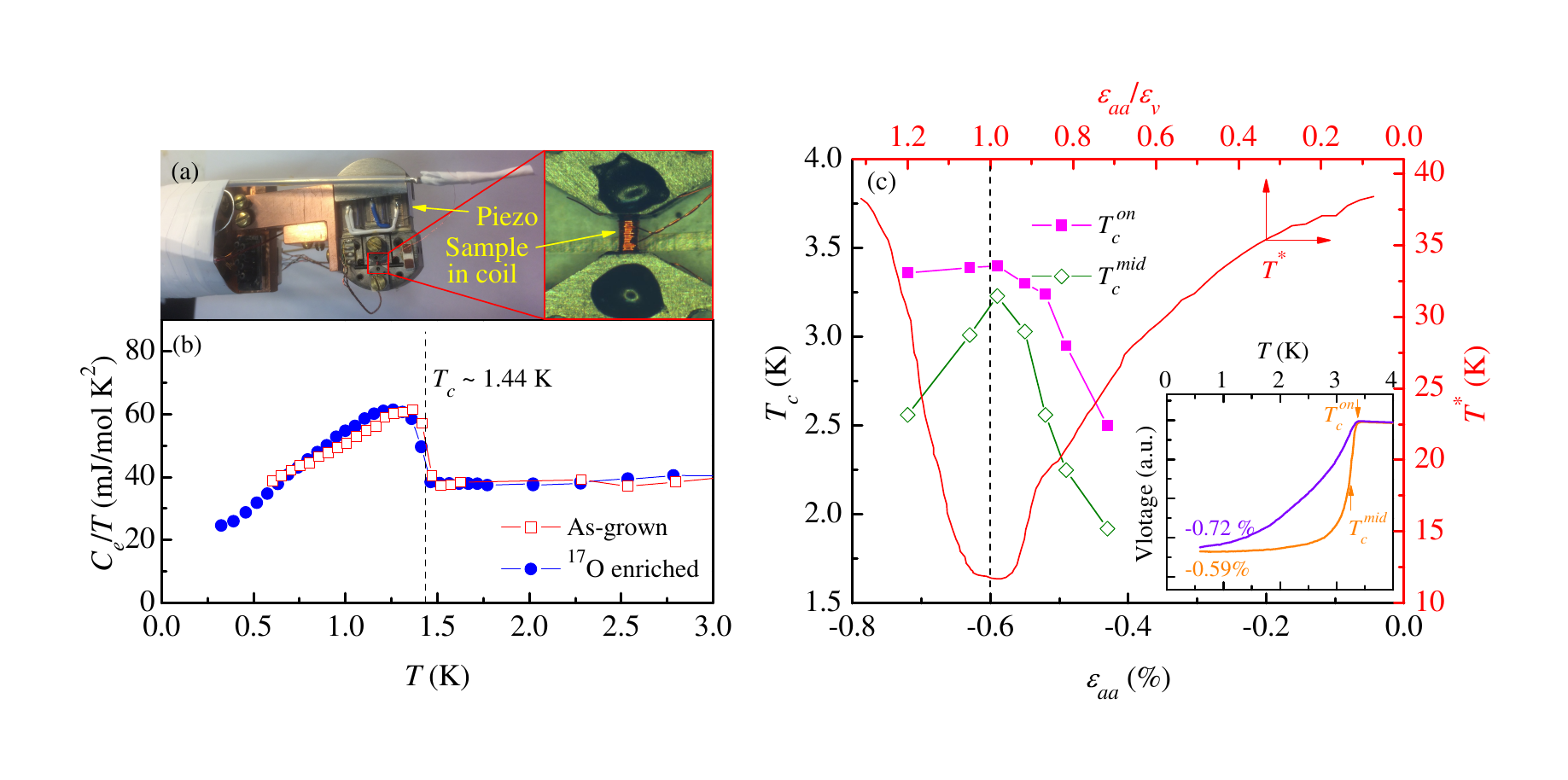}
\vspace*{-20pt}
\caption{\label{Suppfig:Tc} (a) A photograph of the uniaxial stress apparatus. The stress/strain effect is applied through a set of piezoelectric actuators. The forces are applied uniaxially, and the strain response is measured using a capacitive dilatometer. A small coil is made {\it in-situ} surrounding the sample that is bonded between two pairs of titanium plates. (b) Electronic specific heat divided by temperature of the Sr$_2$RuO$_4$ samples before and after $^{17}$O enrichment. Superconducting transitions are clearly visible at about 1.44 K. (c) Strain dependence of superconducting transition from ac magnetization measurements, with critical temperatures $T_c^{on}$ and $T_c^{mid}$ defined in the inset. Maximal $T_c$ is realized near the strain $\varepsilon_{aa}$=$-$0.6\%. The top-right frame displays the profile of $T^*$ reproduced from Ref.~\cite{SBarber:2018}. The strains from the respective experiments are aligned using the respective measured maxima in superconducting critical temperature, $T_c(\varepsilon_{aa}$) \cite{SStrainCalibration}.}
\end{figure*}

Figure \ref{Suppfig:Tc}a is a photograph of the set-up for our NMR measurements under strain. The compressive uniaxial pressure is generated by a set of piezoelectric actuators\cite{SHicks:2014a}. A Sr$_2$RuO$_4$ sample is glued between two pairs of titanium plates with stycast 2850 (black). To get the best filling factor, a small NMR coil (about 25 turns) is made {\it in-situ} surrounding the sample after the stycast hardens with 25 $\mu$m Cu wire.

The quality of the Sr$_2$RuO$_4$ sample measured is characterized by specific heat measurements, as shown in Fig.~\ref{Suppfig:Tc}b. The superconducting transitions are clearly visible in $C_e/T$ before and after annealing in $^{17}$O atmosphere, with $T_c$$\approx$1.44 K essentially unaffected. Here $C_e$ is the electronic contribution to specific heat. The jump in $C_e/T$ at the transition as well as the normal state Sommerfeld coefficient are also in good agreement with previous findings\cite{SDeguchi:2004}. All these measurements guarantee the high quality of the sample studied in this work.

\begin{figure*}[ht]
\vspace*{-20pt}
\includegraphics[width=16cm]{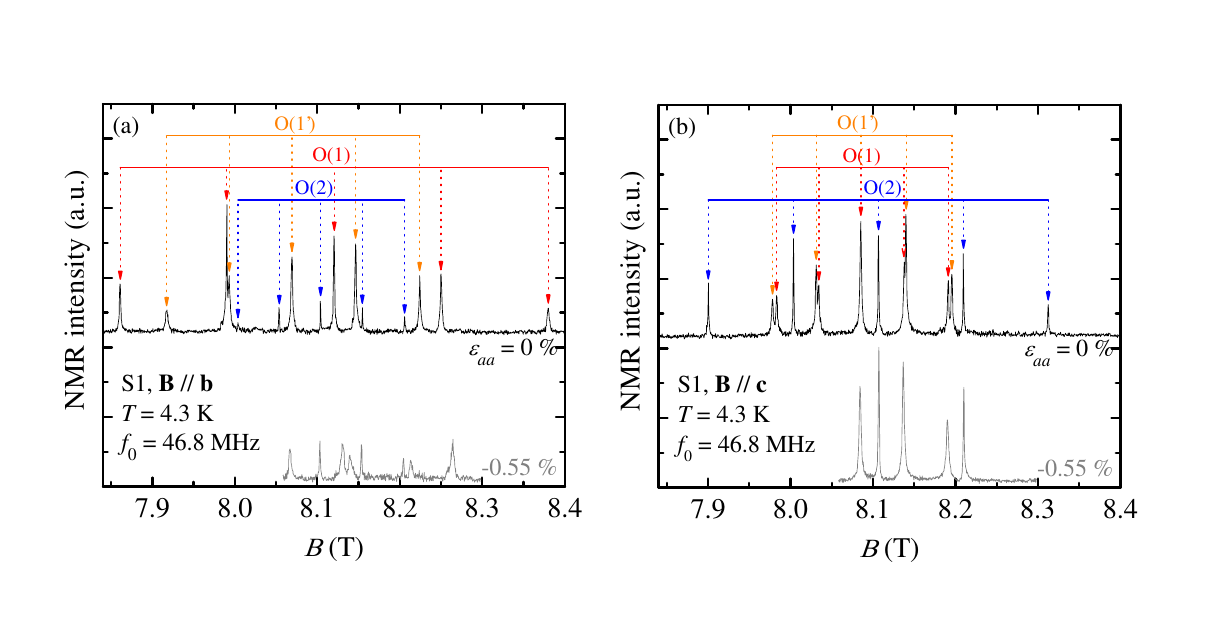}
\vspace*{-20pt}
\caption{\label{Suppfig:FullFieldSweep} Comparison of $^{17}$O NMR spectra of Sr$_2$RuO$_4$ for strain $\varepsilon_{aa}$=0 (black) and $-$0.55\% (grey) where $T_c$ approaches the maximum. (a) measurements with $\mathbf{B}$$\parallel$$\mathbf{b}$, and (b) with $\mathbf{B}$$\parallel$$\mathbf{c}$. At $\varepsilon_{aa}$=0 and $\mathbf{B}$$\parallel$$\mathbf{c}$, the splitting of O(1) and O(1$'$) satellite peaks is due to a small angular misalignment ($\sim$5$\textordmasculine$). }
\end{figure*}

In Fig.~\ref{Suppfig:FullFieldSweep}, we present the field-sweep $^{17}$O NMR spectra of Sr$_2$RuO$_4$ for both $\mathbf{B}$$\parallel$$\mathbf{b}$ (left panel) and $\mathbf{B}$$\parallel$$\mathbf{c}$ (right panel). All the peaks in this field region can be assigned to signals from O(1), O(1$'$) and O(2) sites, and no extra peaks can be identified. This excludes the impurity phases from other members in Sr$_{n+1}$Ru$_n$O$_{3n+1}$ family. For each O site, it shows one central peak ($\frac{1}{2}\leftrightarrow-\frac{1}{2}$) and four satellite peaks corresponding to $\pm\frac{1}{2}\leftrightarrow\pm\frac{3}{2}$ and $\pm\frac{3}{2}\leftrightarrow\pm\frac{5}{2}$, respectively.

\begin{table}[ht]
\tabcolsep 0pt \caption{\label{TabS1} Comparison of Knight shifts and components of the EFG tensor in Sr$_2$RuO$_4$ for $\varepsilon_{aa}$=0 ($T_c$=1.44 K) and $-$0.55\% ($T_c$=3.3 K). Measurements made at 4.3 K. The asymmetry parameter is calculated by $\eta$=($\nu_x$$-$$\nu_y$)/$\nu_z$. The results are from sample S1.}
\vspace*{-0pt}
\begin{center}
\def\temptablewidth{1\columnwidth}
{\rule{\temptablewidth}{1pt}}
\begin{tabular*}{\temptablewidth}{@{\extracolsep{\fill}}cccccc}
 Sites    &               & Quantities               & $\varepsilon_{aa}$=0   & $\varepsilon_{aa}$=$-$0.55\%  & Note                        \\ \hline
  O(1)    & $K_{1}$       & $K_{1\parallel}$ (\%)    &    $-$0.15(1)~         &        $-$0.28(2)~            &                             \\
          &               & $K_{1c}$ (\%)            &    $+$0.29(1)~         &        $+$0.30(1)~            &                             \\
          & EFG(1)        & $\nu_{1a}$ (MHz)         &    $-$0.444(4)         &        $-$0.469(7)            &  $\nu_{1y}$                 \\
          &               & $\nu_{1b}$ (MHz)         &    $+$0.755(5)         &        $+$0.778(9)            &  $\nu_{1z}$=$\nu_{1Q}$      \\
          &               & $\nu_{1c}$ (MHz)         &    $-$0.311(3)         &        $-$0.309(5)            &  $\nu_{1x}$                 \\
          & Asymmetry     & $ \eta_1$                &     ~~0.175(3)         &         ~~0.206(5)            &                             \\ \hline
  O(1$'$) & $K_{1'}$      & $K_{1'\perp}$ (\%)       &    $+$0.48(1)~         &        $+$0.52(2)~            &                             \\
          &               & $K_{1'c}$ (\%)           &    $+$0.29(1)~         &        $+$0.30(1)~            &                             \\
          & EFG(1$'$)     & $\nu_{1'a}$ (MHz)        &    $+$0.759(6)         &        $+$0.730(9)            &  $\nu_{1'z}$=$\nu_{1'Q}$    \\
          &               & $\nu_{1'b}$ (MHz)        &    $-$0.445(4)         &        $-$0.425(6)            &  $\nu_{1'y}$                \\
          &               & $\nu_{1'c}$ (MHz)        &    $-$0.314(4)         &        $-$0.305(5)            &  $\nu_{1'x}$                \\
          & Asymmetry     & $\eta_{1'}$              &     ~~0.172(3)         &         ~~0.164(5)            &                             \\ \hline
  O(2)    & $K_{2}$       & $K_{2b}$ (\%)            &    $+$0.055(6)         &        $+$0.066(9)            &                             \\
          &               & $K_{2c}$ (\%)            &    $+$0.021(5)         &        $+$0.015(3)            &                             \\
          & EFG(2)        & $\nu_{2a}$ (MHz)         &    $-$0.300(2)         &        $-$0.303(3)            &  $\nu_{2y}$                 \\
          &               & $\nu_{2b}$ (MHz)         &    $-$0.300(2)         &        $-$0.299(3)            &  $\nu_{2x}$                 \\
          &               & $\nu_{2c}$ (MHz)         &    $+$0.600(3)         &        $+$0.602(4)            &  $\nu_{2z}$=$\nu_{2Q}$      \\
          & Asymmetry     & $\eta_2$                 &     ~~0.000(1)         &         ~~0.007(2)            &                             \\
\end{tabular*}
{\rule{\temptablewidth}{1pt}}
\end{center}
\end{table}

These satellite peaks arise from nuclear quadrupole interaction with the electric field gradient (EFG) at the nuclear site, as described by
\begin{equation}
H_Q=\frac{eQV_{zz}}{4I(2I-1)}[3\hat{I_z}^2-\mathbf{\hat{I}}^2+\eta(\hat{I_x}^2-\hat{I_y}^2)],\tag{S1}
\label{EqS1}
\end{equation}
where $\mathbf{\hat{I}}$=$(\hat{I_x},\hat{I_y},\hat{I_z})$ is nuclear spin operator, $Q$ is nuclear quadrupole moment, and $\eta$=($V_{xx}$$-$$V_{yy}$)/$V_{zz}$ is the asymmetry parameter with $V_{xx}$, $V_{yy}$ and $V_{zz}$ being the components of the EFG tensor. Here we adopt the convention $V_{zz}$$\geq$$V_{xx}$$\geq$$V_{yy}$, and $V_{xx}$$+$$V_{yy}$$+$$V_{zz}$=0. In Sr$_2$RuO$_4$, $V_{zz}$ is along Ru-O bonding\cite{SMukuda:1998}. This allows us to determine principle-axis nuclear quadrupole resonance (NQR) frequency $\nu_Q$=$\nu_z$ from the spectra shown in Fig.~\ref{Suppfig:FullFieldSweep}. Note that $\nu_z$ is related to $V_{zz}$ by
\begin{equation}
\nu_z=\frac{3eQV_{zz}}{2I(2I-1)h}.\tag{S2}
\label{EqS2}
\end{equation}
Other components of NQR frequencies conform to the formula:
\begin{equation}
\nu_Q'=\nu_Q[\frac{3\cos^2\theta-1}{2}+\frac{\eta}{2}\sin^2\theta\cos2\phi],\tag{S3}
\label{EqS3}
\end{equation}
where $\theta$ and $\phi$ are respectively polar and azimuthal angles as defined in regular $\mathbf{xyz}$-frames, see Fig.~\ref{Suppfig:EFGMeasured}a. Eq.~(\ref{EqS3}) also enables us to verify the sample orientation with respect to magnetic field. In fact, for $\mathbf{B}$$\parallel$$\mathbf{c}$ and $\varepsilon_{aa}$=0, we should expect the NQR peaks of O(1) and O(1$'$) to merge. The splitting of them seen in Fig.~\ref{Suppfig:FullFieldSweep}b is a consequence of small angular misalignment which we estimate to be $\theta$$\sim$5$\textordmasculine$ according to Eq.~(\ref{EqS3}).

Table \ref{TabS1} summarizes all the physical parameters of O(1), O(1$'$) and O(2) sites after the correction of angular misalignment. The results at ambient pressure are in good agreement with that reported by Mukuda {\it et al} \cite{SMukuda:1998}.

\section{\textbf{SM \Rmnum{2}: S\lowercase{train dependent} $\nu_{Q}$ -- \lowercase{experimental and theoretical}}}

\begin{figure*}[ht]
\vspace*{-20pt}
\hspace*{-15pt}
\includegraphics[width=19cm]{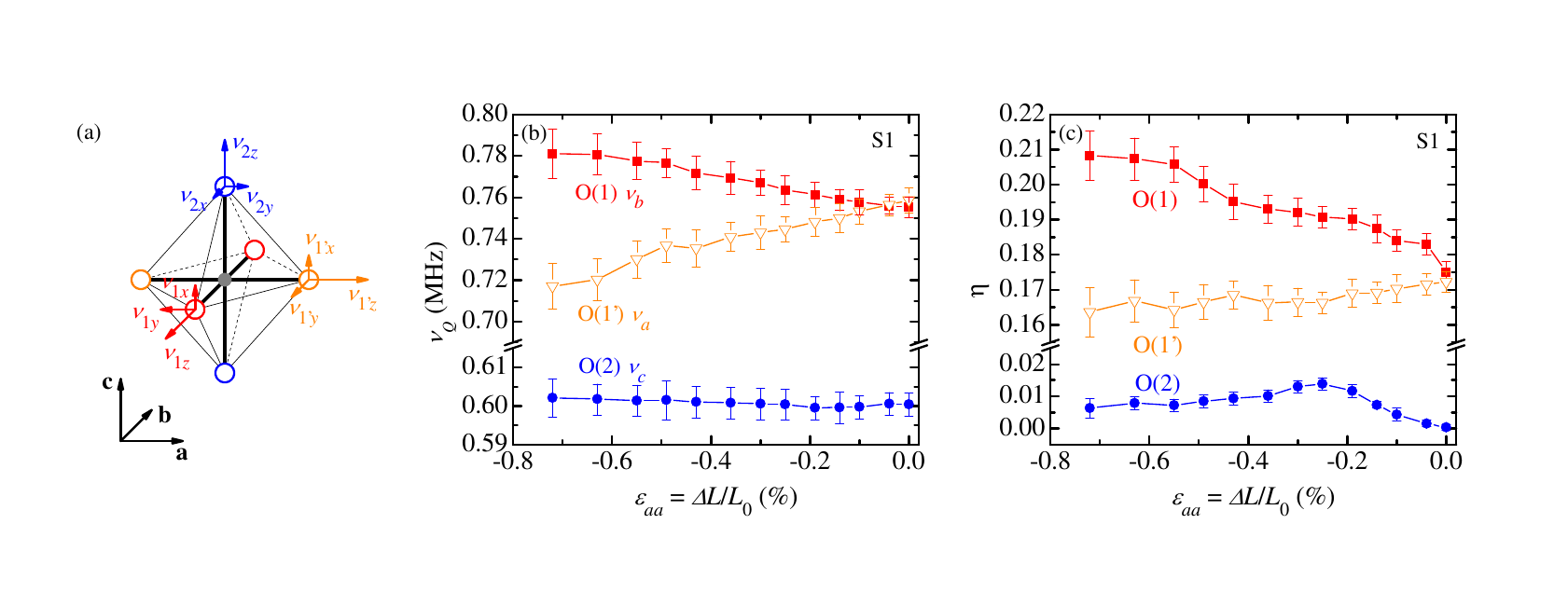}
\vspace*{-30pt}
\caption{\label{Suppfig:EFGMeasured} (a) Schematic sketch of EFG tensors at O sites of Sr$_2$RuO$_4$. The principle component $\nu_z$ is along the Ru-O bond, and the length of the arrows characterizes the magnitude of $\nu_i$ ($i$=$x$,$y$,$z$). (b) and (c) show strain dependence of $\nu_Q$(=$\nu_z$) and asymmetry parameter $\eta$, respectively. }
\end{figure*}

Under strain, the peaks of O(2) sites remain essentially unchanged, while both O(1) and O(1$'$) change drastically. In particular, for $\mathbf{B}$$\parallel$$\mathbf{c}$, the satellite peaks of O(1) and O(1$'$) merge ``coincidentally" when $\varepsilon_{aa}$=$-$0.55\%, implying that the two move at different rates under strain. The strain dependencies of $\nu_Q$ and $\eta$ are displayed in Fig.~\ref{Suppfig:EFGMeasured}b-c. Evidently, the changes of $\nu_{Q}$ in O(1) and O(1$'$) are of opposite signs. This is because an expansive strain is induced along $\mathbf{b}$-axis, {\it i.e.} $\varepsilon_{bb}$$>$0, which is characterized by the Poisson's ratio $-$$\varepsilon_{bb}/\varepsilon_{aa}$=0.40 for Sr$_2$RuO$_4$\cite{SOkuda:2002}. We note that the ratio of the slopes in $\nu_{1Q}(\varepsilon_{aa})$ and $\nu_{1'Q}(\varepsilon_{aa})$ is very close to Poisson's ratio.

Theoretically, $\nu_Q$ usually consists of two contributions: point charge (ionic) of other ions and on-site hole in O $p$ orbitals,
\begin{equation}
\nu_Q=\nu_Q^{ionic}+\nu_Q^{hole},\tag{S4}
\label{EqS4}
\end{equation}
we shall consider them separately. The ionic term can be calculated by (in SI unit):
\begin{equation}
\nu_Q^{ionic} [\text{Hz}]=\frac{1}{4\pi\epsilon_0}\frac{3eQV_{zz}^{ionic}}{2I(2I-1)h}(1-\gamma_{\infty}),\tag{S5}
\label{EqS5}
\end{equation}
where nuclear spin $I$=5/2, quadruple moment $Q$=$-$0.026$\times$10$^{-28}$ m$^2$, and $\gamma_{\infty}$ refers to the Sternheimer antishielding factor which accounts for the contribution from the distortion of the O ion both by the local EFG and by the quadrupolar field of the nucleus\cite{SGarcia:1989}. This antishielding factor turns out to be not important in Sr$_2$RuO$_4$, much weaker than in cuprates\cite{SGarcia:1989}, we therefore ignore it hereafter. $V_{zz}^{ionic}$ can be calculated with the crystalline lattice parameters $a$=$b$=3.8603 \AA, and $c$=12.729 \AA,  and coordinates of the ions Sr$^{2+}$ (0.5, 0.5, 0.1468), Ru$^{4+}$ (0, 0, 0), O(1)$^{2-}$ (0.5, 0, 0) and O(2)$^{2-}$ (0, 0, 0.1619)\cite{SOguchi:1995}.

\begin{figure*}[ht]
\vspace*{-20pt}
\hspace*{-15pt}
\includegraphics[width=19cm]{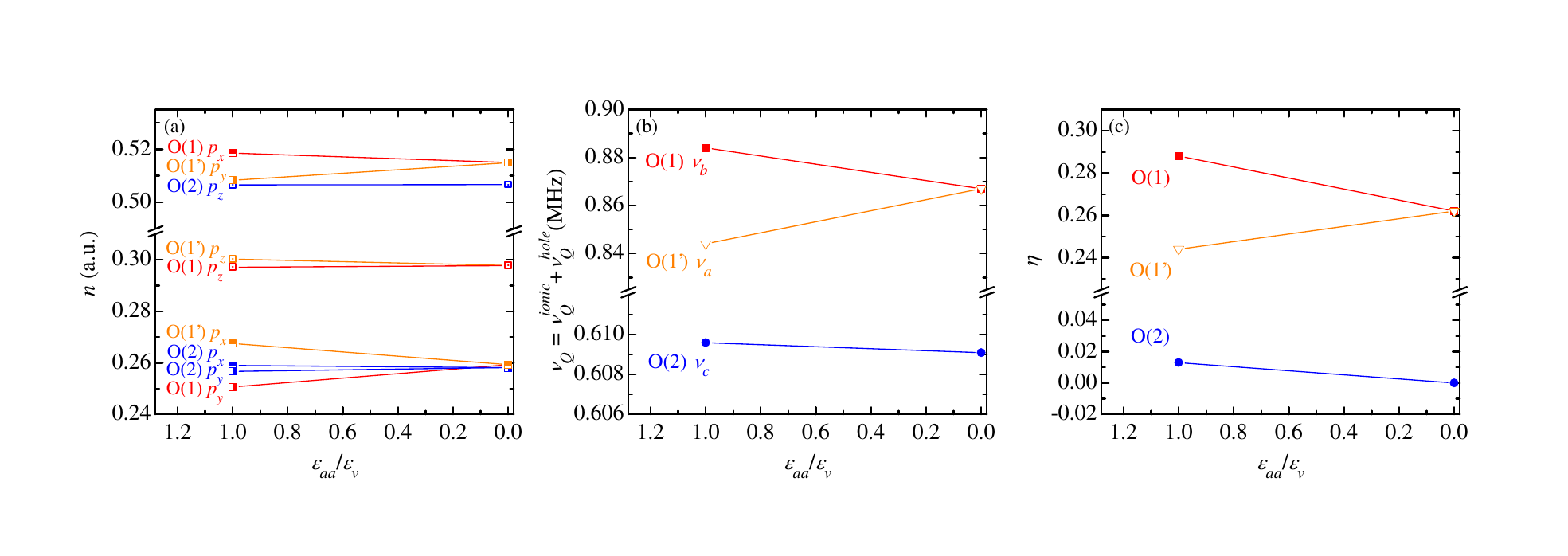}
\vspace*{-30pt}
\caption{\label{Suppfig:EFGCalculated} (a) Comparison of hole content of O orbitals for $\varepsilon_{aa}$=0 and $\varepsilon_{v}$ where the vHS in Sr$_2$RuO$_4$ is realized theoretically. (b) and (c) display calculated quadrupolar frequency ($\nu_Q$) and asymmetry parameter ($\eta$) as a function of $\varepsilon_{aa}$, respectively.}
\end{figure*}

The on-site hole contribution $\nu_Q^{hole}$ is proportional to the hole content ($n$) in each O orbitals, and the latter can be obtained from DFT calculations by integrating the partial density of states up to Fermi energy, viz.
\begin{equation}
2-n=\int_{-\infty}^{E_F} N(E)dE.\tag{S6}
\label{EqS6}
\end{equation}
The variation of $n$ for each O orbitals are displayed in Fig.~\ref{Suppfig:EFGCalculated}.

Taking O(1) $p_x$ orbital as an example, the yielded quadrupoar frequencies are ($\nu_{a}$, $\nu_{b}$, $\nu_{c}$)$_{1,p_x}$=$n_{1,p_x}$($q_{xa}$, $q_{xb}$, $q_{xc}$), where the ratios $q_{xa}$=$-2q_{xb}$=$-2q_{xc}$=2.452 MHz for $^{17}$O according to previous reports on cuprates\cite{SHaase:2004}. The total quadrupolar frequency should be the sum of the contributions from all the three $p$ orbitals for each O site.

The calculated quadrupolar frequencies $\nu_Q$ and the associated asymmetry pramaeter $\eta$ are shown in Fig.~\ref{Suppfig:EFGCalculated}b and c, respectively.

Comparison can be made for $\nu_Q$ and $\eta$ between measured (Fig.~\ref{Suppfig:EFGMeasured}) and calculated (Fig.~\ref{Suppfig:EFGCalculated}) results. Regardless of some difference in magnitude, agreement between experiment and theory in the evolution trend upon strain effect is striking in both $\nu_Q$ and $\eta$.

\section{\textbf{SM \Rmnum{3}: S\lowercase{pin-lattice relaxation rate}}}

\begin{figure}[tbh]
\includegraphics[width=16cm]{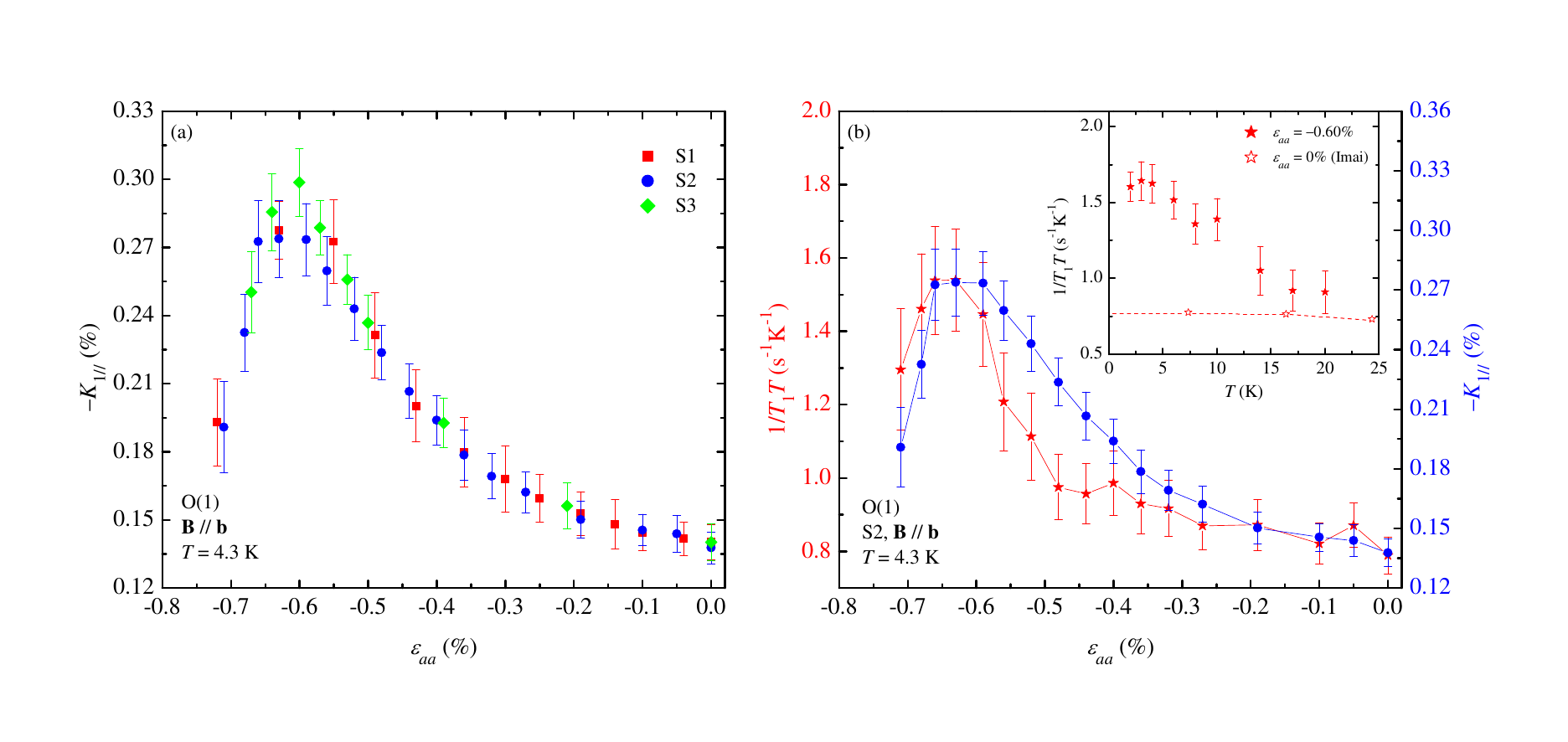}
\vspace*{-15pt}
\caption{(a) Strain dependence of $-$$K_{1\parallel}$ measured from different Sr$_2$RuO$_4$ samples. The inset shows the Korringa ratio $\alpha$$\equiv$$S/K_{1\parallel}^2T_1T$. (b) Magnetization recovery $[T_{1}T]^{-1}$ of central transition for O(1) site as function of strain $\varepsilon_{aa}$, recorded at $T$=4.3 K and with magnetic field aligned with $\mathbf{b}$-axis. The inset shows a variation with temperature.  }%
\label{Suppfig:relaxationStrain}%
\end{figure}

Additional evidence for a vHs comes from the measurements of the spin-lattice relaxation rate $[T_{1}T]^{-1}$ of Sr$_2$RuO$_4$ as shown in Fig.~\ref{Suppfig:relaxationStrain}b. The $[T_{1}T]^{-1}$ is recorded for the central transition of the O(1) site and for field $\mathbf{B}\parallel\mathbf{b}$. As a means to extract the strain dependence of the relaxation rate in a minimum of measurement time, the recovery curves at high ($\varepsilon_{aa}=\varepsilon_v$) and low strain ($\varepsilon_{aa}=0$) were established to follow the appropriate form for spin $I$=5/2, and dominantly magnetic relaxation governing selective irradiation of the central transition. Between these endpoints, a single recovery was recorded, with short delay time selected prior to application of the echo read sequence, so that the relaxation rate could be inferred from the recorded signal amplitude.

As shown in Fig.~\ref{Suppfig:relaxationStrain}b, the relaxation is maximum at the strain where the shifts are extremal, consistent with the vHs-tuning scenario. Although only a narrow temperature range is covered, a temperature dependence is clearly evident in the inset, where the behavior is contrasted to the zero strain results of Ref.~\cite{SMukuda:1998}. The variation could originate partially or entirely from proximity to the vHs, with the remainder related to correlations. Note that the singularity in two dimensions scales as $\ln(t/T)$, with $t$ the relevant hopping integral, and its effect on thermodynamic properties is rapidly diminished due to thermal broadening of the Fermi function.

In order to investigate the magnetic fluctuations of Sr$_2$RuO$_4$, we also consider the Korringa ratio $\alpha$$\equiv$$S_K/(K_s^2T_1T)$, where $S_K$=$(\hbar/4\pi k_B)(\gamma_e/\gamma_n)^2$ with $\gamma_e$=2.8025$\times$10$^4$ MHz/T being electron gyromagnetic ratio. The standard analysis assumes an \textit{isotropic} hyperfine interaction and a single susceptibility. Then, for the case of uncorrelated electrons, $\alpha$=1 \cite{SKorringa:1950}. In the presence of antiferromagnetic correlations, the enhanced $\chi(\mathbf{q})$ around the antiferromagnetic wave vector $\mathbf{q}$ promotes $1/T_1T$ but has little effect on $K_s$, which renders $\alpha$$>$1. The situation is opposite for ferromagnetic correlations, that is, $\alpha$$<$1. A quantitative analysis for \textit{anisotropic} coupling\cite{SHirata-ModifiedKorringa}, as applies here, requires a more detailed angular-dependent study of both shifts and relaxation rates. Consider, for example, the uncorrelated case, and coupling only to the in-plane 2$p_x$ orbital at the O(1) site with $\mathbf{B}\parallel\mathbf{b}$, where the modified $\alpha$$\approx$2.5 is expected. For Sr$_2$RuO$_4$, neutron scattering results indicated antiferromagnetic fluctuations\cite{SSidis:1999} and a broad component at small wavevector\cite{SBraden:2002}. A quantitative interpretation of $\alpha$ is potentially complicated by the multiorbital/multiband nature of Sr$_2$RuO$_4$. Nevertheless, a trend consistent with enhancement of ferromagnetic fluctuations appears in the inset to Fig.~\ref{Suppfig:relaxationStrain}a, where a strong minimum-a reduction of 60\%-in $\alpha$ is centered around $\varepsilon_{aa}$=$\varepsilon_v$, for this orientation of magnetic field.

\section{\textbf{SM \Rmnum{4}: C\lowercase{omments on the $^{17}$\uppercase{O NMR} shifts for $\mathbf{\uppercase{B}}\parallel\mathbf{c}$}}}

In the main text, the NMR Knight shifts of Sr$_2$RuO$_4$ for $\mathbf{B}\parallel\mathbf{c}$ were not closely examined. In part, this is because of an apparently reduced sensitivity to the vHs. In particular, for all strains $|\varepsilon_{aa}|<|\varepsilon_v|$, the central transition for the O(1,1$'$) sites are only weakly changing and remain unresolved, indicating cancellation effects of contributions to the total shifts. Large changes \textit{are} observed for $|\varepsilon_{aa}|>|\varepsilon_v|$, where large drops in spin susceptibility and severe line-broadening are qualitatively consistent with inhomogeneous strain within the measured sample volume and an accompanying amplified sensitivity to the inequivalent environments.

\begin{figure*}[ht]
\vspace*{-20pt}
\hspace*{-15pt}
\includegraphics[width=8cm]{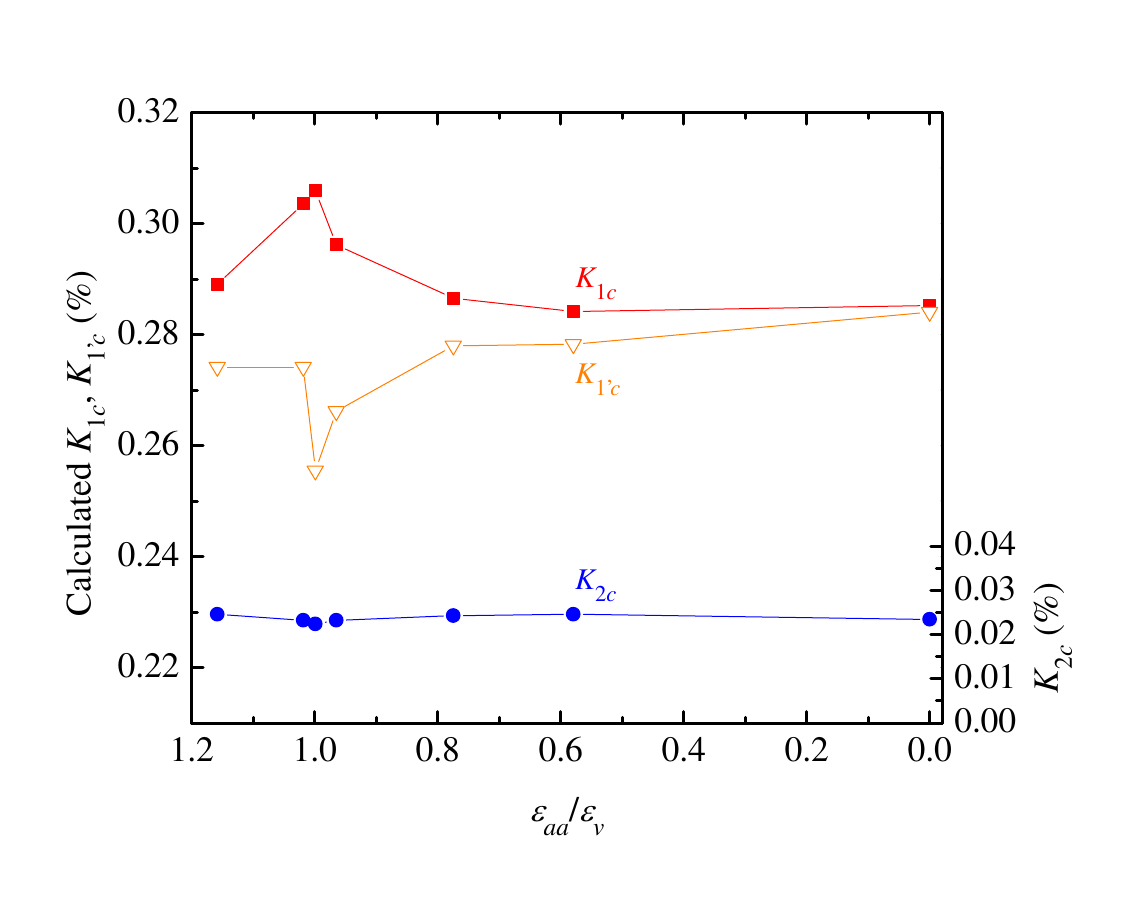}
\vspace*{-15pt}
\caption{\label{Suppfig:CalculatedKnightc} Calculated total Knight shift of Sr$_2$RuO$_4$ for $\mathbf{H}\parallel\mathbf{c}$ for the three sites, O(1), O(1$'$) and O(2). Some discussion of the disparities between these results and what is observed experimentally is included in the main text. }
\end{figure*}

In the DFT calculations for the same quantities, $K_{2c}$ does show essentially full cancellation of the DOS effects, as shown in Fig. \ref{Suppfig:CalculatedKnightc}. On the other hand, both $K_{1c}$ and $K_{1'c}$ appear quite sensitive to the vHs, and, interestingly,
in both Fermi and orbital terms. As discussed in the main text, there are two
mechanisms by which O electrons can acquire an orbital moment: directly induced by the
external field, and via spin-orbit coupling to the induced spin moment. Our calculations
show the former effect in $K_{1c}$ and $K_{1'c}$ to be strong, and {\it opposite in sign to the spin mechanism}. In the raw calculations the amplitude of the orbital shifts is too small
to ensure a full cancellation, but, as discussed in the main text, spin-orbit effects may be
considerably enhanced by correlation (Ref.~\cite{SKim:2018}). Assuming a semiphenomenological
approach, we plot in Fig.~\ref{Suppfig:CalculatedKnightc} the sum of all contributions to $K_{1c}$ and $K_{1'c}$, {\it multiplying the orbital part by a factor of four}, without adding any van Vleck constant. The result still show a small split between $K_{1c}$ and $K_{1'c}$ (albeit smaller than the measured peak widths) and an overall good agreement with the measurements. As we stated in the main text, the NMR spectra for the field parallel to $\mathbf{c}$ require further investigation.


\begin{thebibliography}{54}
\bibitem[{Ann()}]{AnnealNote}
\bibinfo{note}{A post-anneal in air at reduced temperatures (ca. 450
  $^{\circ}$C) led to smaller-than-expected absorption from the apical O(2)
  sites, indicating that the $^{17}$O substitution to the O(1) sites likely
  proceeds through the apical sites.}
\bibitem[{Str({\natexlab{a}})}]{StrainMeasurement}
\bibinfo{note}{The clamping structure is not perfectly rigid, and as a
  consequence, the actual sample compression is smaller than what is measured
  by the dilatometer \cite{Hicks:2014b}.}
\bibitem[{Sto()}]{StonerTensor}
\bibinfo{note}{In principle, the polarization operators and the Stoner factors
  in this formula are matrices in the band indices and reciprocal lattice
  vectors. In the discussion here, we neglect these complexities.}
\end{thebibliography}

\end{document}